\documentclass[12pt]{article}
\usepackage{amssymb,amsmath,amsthm,amsfonts,amscd}
 \usepackage{graphicx}
\newcommand\be{\begin{equation}}
\newcommand\ee{\end{equation}}
\newcommand\bea{\begin{eqnarray}}
\newcommand\eea{\end{eqnarray}}

\newcommand{\fatalpha}{{\bf \alpha \kern -0.44em \alpha}}
\newcommand{\fatsigma}{{\bf \sigma \kern -0.54em \sigma}}
\newcommand{\tpchi}{{\bf \chi \kern -0.35em \chi}}
\newcommand{\llambda}{{\bf \lambda \kern -0.45em \lambda}}

\renewcommand{\theequation}{\arabic{equation}}
\renewcommand{\theequation}{\thesection.\arabic{equation}}
\bibliography{plain}
\title{\bf Floating Entanglement Witness Measure and Genetic Algorithm }\vspace{20mm}
\author{A. Baghbanpourasl$^{a}$ \thanks{E-mail address: baghban@gaalaan.com} ,  G. Najarbashi  $^{b}$
 \thanks{E-mail address: najarbashi@tabrizu.ac.ir} , M. Seyedkazemi $^{c}$ \thanks{E-mail address: mohsen.seyedkazemi@gmail.com}
 \\ $^a${\small Department of Physics, Amir Kabir University, P. O. Box 15875-4413, Tehran, Iran.} \\ $^b${\small Department
of Physics, Mohaghegh Ardabili University, Ardabil 56199-11367,
Iran.} \\ $^c${\small IAU-Ardabil Branch, Young Researchers Club,
Ardabil, Iran}} \pagebreak

 \newtheorem{thm}{Theorem}

 \newtheorem{prop}[thm]{Proposition}
 \newtheorem{defn}[thm]{Definition}
 
\vspace{20mm}
\begin{document}
\maketitle \vspace{15mm}
\newpage
\begin{abstract}
In this paper based on the notion of entanglement witness,  a new
measure of entanglement called floating entanglement witness
measure is introduced which satisfies  some of the usual
properties of a good entanglement measure. By exploiting genetic
algorithm, we  introduce a classical algorithm that computes
floating entanglement witness measure. This algorithm also
provides a  method for finding entanglement witness for a given
entangled state.
\end{abstract}
{\bf AMS: 65C99; 81P99}


 \vspace{70mm}
\newpage
\section{Introduction}
 Entanglement is one of the most interesting properties of
quantum mechanics and the key resource of some quantum information
and quantum computation processes, such as teleportation, dense
coding and quantum key distribution \cite{nielsen1,ekert1}.
However characterizing and  measuring the entanglement has
tantalized physicists since the earliest days of quantum
mechanics, and even today there is no  general qualitative and
quantitative theory of entanglement.
\par
Among the known criterions for distinguishing between separable
and entangled states
\cite{peres1,horod1,wootters1,horod2,lewen1,nielsen2,lewen2},
entanglement witnesses (EWs) have a special importance for
detecting the presence of entanglement. Finding EW that leads to
solving separability problem, is an interesting but
computationally a demanding job as it has been shown
 that the separability problem lies in the class of
NP-hard problems \cite{gurvitz}. There has been a lot of efforts
for finding EW for given quantum state; some of them  propose EW
for some special cases \cite{toth1} while others try to find
approximate solutions by using different methods like semidefinite
and linear programming \cite{doherty1,najar}. In addition to the
question of "Is it entangled?" there is the question of "How much
entangled?". Some measures of entanglement related to EW are
introduced by Bertlemann \emph{et al} \cite{bertl}, Brandao and
Vianna \cite{branda1,branda2}.
\par
The aim of this paper is two fold: firstly by defining a slightly
different definition than of EW which is called floating
entanglement witness (FEW), we introduce a new measure of
entanglement namely FEW measure. Secondly based on the genetic
algorithm (GA) we offer an algorithm which computes FEW measure
and also finds EW for every entangle state. GA is a powerful and
intelligent technique for global optimization,
 adaptation and search problems \cite{mitchell,ramos1,ramos2}. It reveals its power especially when we are dealing
 with highly nonlinear and large search spaces.
  This technique is inspired by natural
 evolution of species which is based on \emph{selection}, \emph{inheritance} and
 \emph{mutation}. Readers are referred to appendix I for an overview of GA.
\par
The paper is organized as follows: In section 2 we briefly recall
the definition of the EW and define our concept of the FEW
measure. In section 3 an algorithm is proposed for finding FEW
measure and EW for a given entangled state. Section 4 is devoted
to some  examples such as: Bell, Werner, mixture of GHZ and W
 and one parameter two-qutrit states. The paper is ended
with a brief conclusion and two appendices.
\section{FEW measure}
As mentioned in the introduction one of the methods for detecting
entanglement is applying  EWs. Let us first recall the definition
of entanglement, separability and EWs. A density matrix $\rho$ is
called \emph{separable} or \emph{unentangled} if there are
positive $p_{i}$'s with $\sum_{i} p_{i}=1$ and product states $|
\psi_{i} \rangle = | \alpha_{i}^{(1)} \rangle | \alpha_{i}^{(2)}
\rangle ...  | \alpha_{i}^{(n)} \rangle$ such that
\begin{equation}\label{}
    \rho=\sum_{i} p_{i} | \psi_{i} \rangle \langle \psi_{i} |
\end{equation}
otherwise it is called \emph{entangled}.
\begin{defn} \label{defW}
A Hermitian operator $\mathcal{W}\in\mathcal{B}(\mathcal{H}_{d_{1}}\otimes
\mathcal{H}_{d_{2}}\ldots\otimes \mathcal{H}_{d_{n}})$ (the Hilbert space of bounded operators)
is called an EW  detecting the
entangled state $\rho_{e}$ if \ $ Tr(\mathcal{W}\rho_{e})<0$ and
$\ Tr(\mathcal{W}\rho_{s})\geq 0 $ for all separable state
$\rho_{s}\in \mathcal{S}$.
\end{defn}
 Therefore, if for state $\rho$ we measure
$Tr(\mathcal{W}\rho)<0$, we can be sure that $\rho$ is entangled. \\
This definition has a clear geometrical meaning. Thus, the set of
states for which $Tr(\mathcal{W}\rho)=0$, is a hyperplane in the
set of all states, that cuts this set into two parts. In the part
in which $Tr(\mathcal{W}\rho)>0$, lies the set of all separable
states. The other part (with $Tr(\mathcal{W}\rho)<0$) is the set
of entangled states detectable by $\mathcal{W}$. From this
geometrical interpretation it follows that for each entangled
state $\rho_{e}$, there exist an entanglement witness  detecting
it. This statement is proved in \cite{horod1}.
\begin{defn}\label{defZ}
A traceless Hermitian operator $\mathcal{Z}$ is called a floating
entanglement  witness (FEW) for detecting the entangled state
$\rho_{e}$ if \ $ Tr(\mathcal{Z}\rho_{e})< \min_{\rho_{s}\in
\mathcal{S}} Tr(\mathcal{Z}\rho_{s})$ where $\mathcal{S}$ is  the
set of separable states.
\end{defn}
It is necessary to note that the existence of $\mathcal{Z}$ comes
from the existence of EW for every $\rho_{e}$. Due to the
convexity of the separable states, the minimum in the above
definition comes from the border of separable states (pure product
states). One can easily show that
\begin{equation}
 \mathcal{W}=\mathcal{Z} - \mu I \quad
\quad \mathrm{where} \quad \mu=\min_{\rho_{s}\in \mathcal{S}}
Tr(\mathcal{Z}\rho_{s}),
\end{equation}
is an EW since it satisfies both conditions of EW, i.e.: $
Tr(\mathcal{W}\rho_{s})= Tr(\mathcal{Z}\rho_{s})-\mu \geq 0 $ for
all separable  $\rho_{s}$ and   $
Tr(\mathcal{W}\rho_{_{e}})=Tr(\mathcal{Z}\rho_{_{e}})-\mu<
Tr(\mathcal{Z}\rho_{s})-\mu \leq 0$, which means that
$Tr(\mathcal{W}\rho_{_{e}})<0 $.
\par
A closely related problem to
the EWs is characterization or quantification of entanglement by
EWs (see \cite{branda2}). Based on the concept of FEW we introduce
a new computable FEW measure for quantifying entanglement of a
given quantum state:
\begin{equation} \label{defE}
E(\rho):=  \max\left\{0\ , \ \max_{ \mathcal{Z}\in \mathcal{A}} \big[\min_{\rho_{s}\in \mathcal{S}}
Tr(\mathcal{Z}\rho_{s})-Tr(\mathcal{Z}\rho)\big]\right\},
\end{equation}
where
$$\mathcal{A}:=\left\{\mathcal{Z}\in\mathcal{B}(\mathcal{H}_{d_{1}}\otimes
\mathcal{H}_{d_{2}}\ldots\otimes \mathcal{H}_{d_{n}})\ \ | \ Tr(\mathcal{Z})=0,\ \|\mathcal{Z}\|_{2}
=1 \right\},
$$
and $\|\mathcal{Z}\|_{2} :=\sqrt{Tr(\mathcal{Z}^\dagger
\mathcal{Z})}$, denotes  the Hilbert-Schmidt norm.
 The FEW measure fulfills the following  usual
requirements of an entanglement measure \cite{vedral}:
\begin{prop}\label{defMeas}
$E(\rho)$  satisfies the following properties:
\end{prop}
\begin{description}
    \item[(i)] For every separable state $\sigma_{s}\in \mathcal{S}$, $E(\sigma_{s})=0$.
    \item[(ii)] Local unitary operations leave $E(\rho)$ invariant, i.e,
    \begin{equation}\label{secprop}
    E(\rho)=
    E(U_{1}^\dagger \otimes\ldots\otimes U_{n}^\dagger\rho U_{1}\otimes\ldots\otimes U_{n}).
\end{equation}
    \item[(iii)]FEW measure dose not increase under  local operations and classical communication
(LOCC) protocols \cite{gisin1}, i.e.,
\begin{equation}\label{}
    E\big(\mathcal{E}(\rho)\big)\leq  E(\rho).
\end{equation}
   \item[(iv)]The FEW measure is a convex function, i.e.,
\begin{equation}\label{}
    E\big(\lambda\rho+(1-\lambda)\sigma)\big)\leq
    \lambda E(\rho)+(1-\lambda)E(\sigma).
\end{equation}
\item[(v)] FEW measure is a continuous function.
\end{description}
\emph{Proof:} (i) It is easy to see  that for every $\mathcal{Z}$ we have $\min_{\rho_{s}}
Tr(\mathcal{Z}\rho_{s})-Tr(\mathcal{Z}\sigma_{s})\leq0$  hence $E(\sigma_{s})=0$.\\
(ii) (\ref{secprop}) follows from the invariance
of $\mathcal{S}$ and $\|.\|_{2}$ under local unitary operations.\\
(iii) Although  we did not find rigorous proof for monotonicity of
FEW measure under general LOCC maps $\mathcal{E}$, it can be
proved
 for isometry ones for which we require $\mathcal{E}^\dag\mathcal{E}=\mathcal{E}\mathcal{E}^\dag=I$.
 To see this we note that
$$
 E\big(\mathcal{E}(\rho)\big)=  \max\left\{0\ , \ \max_{ \mathcal{Z}\in \mathcal{A}} \big[\min_{\rho_{s}\in \mathcal{S}}
Tr(\mathcal{Z}\rho_{s})-Tr\big(\mathcal{Z}\mathcal{E}(\rho)\big)\big]\right\}\quad\quad (\mathrm{by \ definition})
$$
$$
= \max\left\{0\ , \ \max_{\mathcal{Z}\in \mathcal{A}} \big[\min_{\mathcal{E}^\dagger(\rho_{s})
\in \mathcal{S}}Tr\big(\mathcal{E}^\dagger(\mathcal{Z})\mathcal{E}^\dagger(\rho_{s})\big)
-Tr\big(\mathcal{E}^\dag(\mathcal{Z})\rho\big)\big]\right\}\quad\quad (\mathrm{isometry})
$$
$$
=\max\left\{0\ , \ \max_{\mathcal{Z}'\in \mathcal{E}^\dagger(\mathcal{A})} \big[\min_{\rho_{s}
\in \mathcal{S}}Tr\big(\mathcal{Z}'\rho_{s}\big)-Tr\big(\mathcal{Z}'\rho\big)\big]\right\}\quad\quad
(\mathcal{E}^\dagger(\mathcal{Z})\rightarrow \mathcal{Z}' )
$$
$$
\leq \max\left\{0\ , \ \max_{ \mathcal{Z}\in \mathcal{A}} \big[\min_{\rho_{s}\in \mathcal{S}}
Tr(\mathcal{Z}\rho_{s})-Tr(\mathcal{Z}\rho)\big]\right\}=E(\rho)
\quad\quad(\mathcal{E}^\dagger(\mathcal{A})\subseteq \mathcal{A} )
$$
where in the second equality we use the fact that
$\mathcal{E}^\dag(\mathcal{S})= \mathcal{S}$, \cite{setinvar} and
$Tr\big(\mathcal{Z}\mathcal{E}(\rho)\big)=Tr\big(\mathcal{E}^\dag(\mathcal{Z})\rho\big)$.\\
(iv) To prove the convexity we note that
$$
E\big(\lambda\rho+(1-\lambda)\sigma)\big)
=\max\left\{0\ , \ \max_{ \mathcal{Z}\in \mathcal{A}} \big[\min_{\rho_{s}\in \mathcal{S}}
Tr(\mathcal{Z}\rho_{s})-Tr\big(\mathcal{Z}(\lambda\rho+(1-\lambda)\sigma\big)\big]\right\}
$$
$$
=\max\left\{0\ , \ \max_{ \mathcal{Z}\in \mathcal{A}} \big[\lambda\min_{\rho_{s}\in \mathcal{S}}
Tr(\mathcal{Z}\rho_{s})+(1-\lambda)\min_{\rho_{s}\in \mathcal{S}}
Tr(\mathcal{Z}\rho_{s})-\lambda Tr\big(\mathcal{Z}\rho\big)
-(1-\lambda)Tr\big(\mathcal{Z}\sigma\big)\big]\right\}
$$
$$
\leq \lambda\max\left\{0\ ,\ \max_{ \mathcal{Z}\in \mathcal{A}}\big[\min_{\rho_{s}\in \mathcal{S}}
Tr(\mathcal{Z}\rho_{s})- Tr(\mathcal{Z}\rho)\big]\right\}+
(1-\lambda)\max\left\{0\ ,\ \max_{ \mathcal{Z}\in \mathcal{A}}\big[\min_{\rho_{s}\in \mathcal{S}}
Tr(\mathcal{Z}\rho_{s})- Tr(\mathcal{Z}\sigma)\big]\right\}
$$
$$
= \lambda E(\rho)+(1-\lambda)E(\sigma).
$$
where the inequality comes from the fact that,
$\max_{\Omega}(f_{1}+f_{2})\leq\max_{\Omega} f_{1}+\max_{\Omega}
f_{1}$, for every function $f_{1}, f_{2}$ defined on a given  set
$\Omega$.\\
(v) To prove continuity, for entangled states $\rho$ and $\sigma$
we suppose  that $E(\rho)>E(\sigma)$, without loss of generality.
By definition
$$
E(\rho)=\max_{ \mathcal{Z}\in \mathcal{A}}\big[\min_{\rho_{s}\in
\mathcal{S}} Tr(\mathcal{Z}\rho_{s})- Tr(\mathcal{Z}\rho) \big]=
Tr(\mathcal{Z}_{\rho}\rho_{s_{0}})- Tr(\mathcal{Z}_{\rho}\rho),
$$
where  $\mathcal{Z}_{\rho}$ and $\rho_{s_{0}}$ are the
$\mathcal{Z}\in \mathcal{A}$ and $\rho_{s}\in \mathcal{S}$ for
which the maximum  occurs in the above equation. On the other hand
we have
$$
E(\sigma)=\max_{ \mathcal{Z}\in
\mathcal{A}}\big[\min_{\sigma_{s}\in \mathcal{S}}
Tr(\mathcal{Z}\sigma_{s})- Tr(\mathcal{Z}\sigma) \big]\geq \forall
\mathcal{Z}  \  \big[\min_{\sigma_{s}\in \mathcal{S}}
Tr(\mathcal{Z}\sigma_{s})- Tr(\mathcal{Z}\sigma)\big]
$$
$$
\Longrightarrow E(\sigma) \geq Tr(\mathcal{Z}_{\rho}\rho_{s_{0}})-
Tr(\mathcal{Z}_{\rho}\sigma)
$$
Therefore
$$
E(\rho)-E(\sigma) \leq  - Tr(\mathcal{Z}_{\rho}\rho)+
Tr(\mathcal{Z}_{\rho}\sigma)=Tr(\mathcal{Z}_{\rho}
(\sigma-\rho))\leq \| \mathcal{Z}_{\rho}
\|_{2}\|\sigma-\rho\|_{2}= \|\sigma-\rho\|_{2}=\epsilon
$$
for some real number $\epsilon \geq 0$. In the last inequality we
have used Cauchy-Schwartz inequality for Hilbert-Schmidt distance.
\par
Another usual property of every measure of entanglement is the
\emph{additivity} $ E(\rho^{\otimes n}) = nE(\rho)$ or
\emph{subadditivity}  $ E(\rho \otimes \sigma) \leq
E(\rho)+E(\sigma)$ problem which for FEW measure remains open for
debate.
\section{Finding FEW measure and EWs using GA}
This section deals with application of GA in finding FEW measure
and EWs. At the first stage the problem is finding a FEW for a
given $\rho$. The fitness function is defined as
\begin{equation}
\mathcal{F}(\rho,\mathcal{Z}):= \min_{\rho_{_{s}}} \
Tr(\mathcal{Z}\rho_{s})-Tr(\mathcal{Z}\rho).
\end{equation}
The task is to  maximize $\mathcal{F}$ over $\mathcal{Z}$ by using
the GA, which implies that GA tends to make $Tr(\mathcal{Z}\rho)$
smaller than $ \min Tr(\mathcal{Z}\rho_{s}) $,
  as well as it increases the difference between these two terms in each chromosome.
  As it is clear from the fitness function we need a
  second optimization procedure that finds $\min \ Tr(\mathcal{Z}\rho_{s}) $, for
  each chromosome. In this subprogram, quasi-Newton (QN) optimization method is
  used by employing optimization function UMINF of IMSL math
  library. For this reason we calculate
  $\mathrm{Tr}(\rho_{s}\mathcal{Z})$ for a large number, $N_{_{QN1}}$, of random
  $\rho_{s}$'s then use  $N_{_{QN2}}$  of them which have
  smaller $\mathrm{Tr}(\rho_{s}\mathcal{Z})$ as initial points for running QN.
  The smallest resulting value is chosen as $\min_{\rho_{_{s}}} \
  Tr(\mathcal{Z}\rho_{s})$.
\par
 Regarding the above
considerations the algorithm goes as follows:
\begin{enumerate}
    \item Read input density matrix $\rho$.
    \item Populate an initial pool of random
$\mathcal{Z}$'s (chromosomes), i.e. produce $\mathcal{Z}$'s with
random parameters.
    \item \label{eval} \begin{enumerate}
        \item \label{findEnt}  Find  $f_{1}:= \min_{\rho_{s}} \ Tr(\rho_{s}\mathcal{Z})$
for each $\mathcal{Z}$ by using QN.
  \item \label{findSep} Find $f_{2}:=Tr(\rho \mathcal{Z})$ for every $\mathcal{Z}$ in the pool.
        \item  Compute the
fitness function $\mathcal{F}=f_{1}-f_{2}$, for every
$\mathcal{Z}$.
    \end{enumerate}
    \item Produce the new generation by doing selection, crossover and
    mutation.
    \item Go to step (\ref{eval}) until the stop
criteria is met, i.e., maximum number of iterations is run.
\item \begin{enumerate} \item Select the chromosome ($\mathcal{Z}$), with maximum $\mathcal{F}$
\item If $\mathcal{F}>0$  then $\mathcal{F}$ is FEW measure and $\mathcal{Z}$ is
FEW. \item Compute $ \mathcal{W}=\mathcal{Z} -\mu I$
$(\mathrm{with}\ \mu=f_{1})$ as an EW for detecting $\rho$.
\end{enumerate}
\item if $\mathcal{F}\leq 0$, $\rho$ is separable.
\end{enumerate}
\par
It is useful to mention that even if we just want to find EW, it
is better to find it through FEW. An EW, $\mathcal{W}$, must
satisfy the conditions in definition \ref{defW}, i.e. a
$\mathcal{W}$ which \ $ Tr(\mathcal{W}\rho)<0 $ for the given
$\rho $, subjected to the constraint $ Tr(\mathcal{W}\rho_{s})\geq
0 $ for all separable states $ \rho_{s}$'s. The latter condition
puts our problem in the field of constrained problems. GA is
adaptable to the constrained problems by defining a proper fitness
function. But the constraint of the definition \ref{defW} slows
down the algorithm and therefore instead of it we use definition
\ref{defZ} which breaks the constraint and greatly speeds up the
algorithm.
\par
 The above algorithm  can find a FEW, an EW  and FEW measure  for any
arbitrary entangled density matrix. The robustness of the
algorithm will be clear in the next section with various examples.
\section{Some  examples }
In this section we discuss some interesting  cases which can help
to clarify the  subject. In all cases we use the fact that any
traceless Hermitian operator acting on the Hilbert space
$\mathcal{H}_{d_{1}}\otimes \mathcal{H}_{d_{2}}...\otimes
\mathcal{H}_{d_{n}}$ can be expressed by identity operator
$I_{d_{1}d_{2}...d_{n}}$ and generators of Lie algebra
$su(d_{i})$, $\lambda^{(d_{i})}_{j}$, as
\begin{equation}
\mathcal{Z}=\sum_{i_{1}=0}^{d_{1}^2-1}\sum_{i_{2}=0}^{d_{2}^2-1}...\sum_{i_{n}=0}^{d_{n}^2-1}
\tau_{i_{1}i_{2}...i_{n}}
\lambda_{i_{1}}^{(d_{1})}\otimes\lambda_{i_{2}}^{(d_{2})}\otimes...\otimes\lambda_{i_{n}}^{(d_{n})},
\quad \tau_{00...0} = 0,
\end{equation}
where $\lambda_{0}^{(d_{i})}=I_{d_{i}}$ and
$\tau_{i_{1}i_{2}...i_{n}}\in \mathbb{R}$.
\par
Typically in GAs, there is no strict rule for choosing parameters
of them. Probability of crossover $P_{c}$ is usually chosen 0.7
and probability of mutation $P_{m}$, is a very small number in
$10^{-3}$ order. Number of population $N_{_{GA}}$ is usually
several hundreds to several thousands.
\par
 In the following examples, crossover is two-point, $P_{c}$ is 0.7, $P_{m}$ is 0.007 and $N_{_{GA}}$ is chosen
   20 (for 2 qubit states) and 10 (for other states)  times bigger than  the number of parameters of
the problem. Convergence of  GA  in our problem starts, depending
on the quantum state, from 20 iteration to higher ones. Therefore
the maximum number of generations $G_{_{GA}}$, in most examples is
chosen 300 which is big enough for finding proper solutions. Also,
there are no lower and upper limits for parameters of QN except
that choosing very small numbers make QN less successful in
finding global minimum and very big numbers make the algorithm
very slow. In this program, $N_{QN1}$  is chosen about 100 times
bigger than the number of parameters of the general form of pure
separable states and $N_{QN2}$ is chosen about this number of
parameters.
\subsection{Two-qubit systems}
It is important both theoretically and experimentally to study
entanglement of qubit systems  and to provide EWs to verify that
in a given state, entanglement is really present.
\par
Here we find  an EW for a given two-qubit density matrix. The most
general form of a traceless Hermitian operator in the space of
two-qubit states can be written as:
\begin{equation}\label{towwit}
\mathcal{Z}=\sum_{i,j=0}^3 \tau_{ij}\sigma_{i}\otimes \sigma_{j},
\quad \tau_{00} =0,
\end{equation}
where $\tau_{ij}\in \mathbb{R}$, $\sigma_{0}=I_{2}$ and
$\sigma_{i}$'s with $i=1,2,3$, are usual Pauli matrices. The
equation (\ref{towwit}) has 15 parameters ($\tau_{ij}$'s) which
have to be determined by GA. The range of $\tau_{ij}$'s is
determined by the condition $\|\mathcal{Z}\|_{2}=1$ in  every
example. Each of the $\tau_{ij}$'s is encoded in a 15 bit binary
number. Therefore every $\mathcal{Z}$ can be encoded in a
chromosome of 225 bits. Parameters of each chromosome are passed
to the QN. In the QN, for finding $f_{1}$, it is sufficient to
take this minimum over the $\rho_{s}$'s in the form $\rho_{s}= |
\phi_{1} \rangle \langle \phi_{1} |\otimes | \phi_{2} \rangle
\langle \phi_{2} | $ which can be parameterized as:
\begin{equation}\label{prod}
  |\phi_{j}
\rangle=\cos(\beta_{j})|0\rangle+e^{i\alpha_{j}}\sin(\beta_{j})|1\rangle,\quad
j=1,2
\end{equation}
where $ \alpha_{j}\in[0,2\pi]$ and $
\beta_{j}\in[0,\frac{\pi}{2}]$. QN algorithm minimizes $
Tr(\mathcal{Z}\rho_{s})$ by determining these four parameters.
\subsubsection{Bell states}
We begin with the simplest case that is finding  an EW for  pure
 Bell state $|\psi_{00}\rangle
\langle \psi_{00} |$,  where
$|\psi_{00}\rangle=\frac{1}{\sqrt{2}}(|00\rangle+|11\rangle)$.
 Crossover is two-point, probability of crossover ($P_{c}$) is 0.7, probability of mutation
($P_{m}$) is 0.007 and numbers of population and generations for
GA are $N_{_{GA}}=350$ and $G_{_{GA}}=80$ respectively and
parameters of QN are $N_{QN1}=400$ and $N_{QN2}=5$. The program
gives us the following solution:
$$\mathcal{W}=\small{\left( \begin{array}{cccc}
 0 &           0.296 &   0.280 &  -0.289+0.001i \\
   0.296 &   0.575 &    0.288-0.001i &   0.283 \\
   0.280 &  0.288+0.001i  &  0.578   &  0.292 \\
  -0.289-0.001i &  0.283 &   0.292   & 0 \\
\end{array}
\right)} $$ with $E(\rho)=0.577$. If we don't impose the condition
$\|\mathcal{Z}\|_{2}=1$, we have:
$$
\mathcal{W}=\small{\left( \begin{array}{cccc}
0   &        0.003i & 0.002i & -1 +
  0.014i \\
  -0.003i &  0.999      &          0 &
  0.002i \\
   -0.002i    &  0 &  0.999    &
 0\\
  -1-0.014i &  -0.002i  &0  &
  0 \\
\end{array}
\right)} $$
 This  is very similar to the EW corresponding to \emph{reduction map}  \cite{hall1}:
\begin{equation}\label{redu}
   \mathcal{W}_{red}=I-2| \psi_{00}\rangle \langle
\psi_{00} |=\left(%
\begin{array}{cccc}
  0  &      0      &   0   & -1\\
         0   & 1  &      0     &    0 \\
         0 &        0  &  1    &     0\\
   -1   &      0   &      0 &   0\\
\end{array}%
\right)
\end{equation}
Of course choosing any Bell state
\begin{equation}\label{bell}
|\psi_{ij}\rangle:=\sigma_{z}^i\otimes\sigma_{x}^j|\psi_{00}\rangle,
\quad i,j=0,1,
\end{equation}
the GA yields an EW similar to reduction EW,
$\mathcal{W}_{red}=I-2| \psi_{ij}\rangle \langle \psi_{ij} |$.
\subsubsection{Werner states}
One of the most important degraded Bell states is Werner state
\cite{werner}. A Werner state in $2 \otimes 2$ system takes the
following form:
\begin{equation}
\rho_{_{W}}= \mathrm{F}| \psi_{00} \rangle \langle \psi_{00} | +
\frac{1-\mathrm{F}}{3}(| \psi_{10} \rangle \langle \psi_{10} |+|
\psi_{01} \rangle \langle \psi_{01} |+ | \psi_{11} \rangle \langle
\psi_{11} |)\quad,\quad 0\leq F\leq1
\end{equation}
where $| \psi_{ij}\rangle$'s are Bell states defined in Eq.
(\ref{bell}). The Werner state $\rho_{_{W}}$ is characterized by a
single real parameter F called fidelity. This quantity measures
the overlap of Werner state with a Bell state $| \psi_{00}
\rangle$. The $\rho_{_{W}}$ is separable for $0\leq
F\leq\frac{1}{2}$ and is entangled for $\frac{1}{2}< F \leq1$. In
these cases which the EW is complicated to write down, we give
only the FEW measure (see Fig. 1).
\subsection{Three-qubit systems}
The procedure for three-qubit states is similar to the one
described in the previous subsection. It is clear that every
three-qubit traceless Hermitian operator $\mathcal{Z}$ can be
written by tensor product of Pauli operators as
\begin{equation}
\mathcal{Z}=\sum_{i,j,k=0}^4 \tau_{ijk}\sigma_{i}\otimes
\sigma_{j}\otimes \sigma_{k}, \quad  \tau_{000}=0
\end{equation}
where the number of parameters $\tau_{ijk}\in \mathbb{R}$, is 63,
so every $\mathcal{Z}$ can be encoded in a string of $15 \times
63=945$ bits. Similar to two-qubit case,  we work with pure
separable states
 $\rho_{s}=| \phi_{1} \rangle \langle \phi_{1} |\otimes
| \phi_{2} \rangle \langle \phi_{2} |\otimes | \phi_{3} \rangle
\langle \phi_{3} |$, for finding $\min_{\rho_{s}} Tr(\mathcal{Z}
\rho_{s})$, where $ | \phi_{j} \rangle $ is defined the same as
Eq. (\ref{prod}). Therefore we have 6 parameters for QN.
\par
For example consider the mixture of $| \mathrm{W} \rangle=(|100
\rangle+ |010\rangle + |001\rangle)/ \sqrt{3}$ and $| \mathrm{GHZ}
\rangle= (|000 \rangle+ | 111 \rangle)/ \sqrt{2}$:
\begin{equation}
\rho_{q}=q | \mathrm{GHZ} \rangle \langle \mathrm{GHZ} | + (1-q)
|\mathrm{W} \rangle \langle \mathrm{W} |, \quad 0 \leq q \leq 1
\end{equation}
 Results for FEW measure which are shown in Fig. 2 are
in agreement with \emph{witnessed entanglement} \cite{branda2}.
\subsection{One parameter two-qutrit state}
The description of qutrit systems is very similar to the one for
qubits. For two-qutrit states the operator $\mathcal{Z}$ can be
expressed as $ \mathcal{Z}=\sum_{i,j=0}^8
\tau_{ij}\lambda_{i}\otimes \lambda_{j}$, $\tau_{00}=0 $, where
$\lambda_{i}$'s with $i=1,...,8$ are the well-known Gell-Mann
matrices (see appendix II). The number of parameters $\tau_{ij}\in
\mathbb{R}$, is 80, so every $\mathcal{Z}$ could be encoded in a
string of $15 \times 80=1200$ bits. We work with pure separable
states  $\rho_{_{s}}=| \phi_{1} \rangle \langle \phi_{1} |\otimes
| \phi_{2} \rangle \langle \phi_{2} |$, for finding $\min
Tr(\mathcal{Z} \rho_{s})$, where
$$
|\varphi_{j}\rangle=e^{i\eta_{j}}\sin(\theta_{j})\sin(\phi_{j})|0
\rangle +e^{i\xi_{j}}\sin(\theta_{j})\cos(\phi_{j})|1
\rangle+\cos(\theta_{j})|2 \rangle, \quad j=1,2
$$
and $\eta_{j},\xi_{j}\in[0,2\pi]$  and
$\theta_{j},\phi_{j}\in[0,\frac{\pi}{2}]$. As last example
consider the one parameter two-qutrit density matrix
\begin{equation}
 \rho_{_{\alpha}}=\frac{2}{7}|\phi_{_{+}}\rangle \langle \phi_{_{+}}| +
\frac{\alpha}{7}\sigma_{+}+ \frac{5-\alpha}{7}\sigma_{-}, \quad 2
\leq \alpha \leq 5
\end{equation}
 where
\begin{equation}
\begin{array}{c}
  |\phi_{_{+}}\rangle=\frac{1}{\sqrt{3}}(|00\rangle+|11\rangle+|22\rangle) \\
  \sigma_{+}=\frac{1}{3}\big(|01\rangle \langle 01| + |12\rangle \langle
12|+ |20 \rangle \langle 20|\big) \\
  \sigma_{-}=\frac{1}{3}\big(|10\rangle \langle 10| + |21\rangle \langle
21|+ |02 \rangle \langle 02|\big)  \\
\end{array}
\end{equation}
It has been shown that $\rho_{_{\alpha}}$ is separable if
$2\leq\alpha\leq3$ and  entangled  if $3<\alpha\leq5$,
\cite{horod2}. FEW measure for these states is shown in Fig. 3.
\section{Conclusion}
We have introduced a new measure of entanglement called FEW
measure, to quantify entanglement of quantum states which is based
 on a slightly different definition than of EW. Some  properties of every measure of entanglement were proved
for this measure. Using GA we proposed an algorithm to compute FEW
 measure for multipartite systems.
 For several examples including two-qubit, three-qubit and two-qutrit systems, the results of this
method were illustrated. As in the examples, this method is
capable of finding numerically these operators and quantity for
every quantum state with any number of dimensions and parties.
This algorithm also provided a method for finding EW for any
entangled state.

 \vspace{1cm}
\setcounter{section}{0}
 \setcounter{equation}{0}
 \renewcommand{\theequation}{I-\arabic{equation}}

{\Large{Appendix I}}\\
\emph{GA}: GA is a search algorithm, based on natural selection
and genetics \cite{mitchell}. In every GA there is a population of
individuals, named chromosomes which a possible solution is
encoded in each of them. A number is assigned to every chromosome
showing its fitness to survive and reproduce. This number is
calculated by fitness function. To form a new generation,
chromosomes are selected based on their fitness, i.e., the fitter
chromosome has more chance to be selected for reproduction.
Tournament selection is one of the selection methods which selects
the fittest chromosome in a number of randomly selected
chromosomes. Selected chromosomes (parents) are subjected to
genetic operations of crossover and mutation to reproduce
offsprings. Crossover is combination of two parents with a
probability, $P_{c}$, creating one or two new chromosomes. Parents
interchange some parts of their chromosomes at some randomly
chosen places, e.g., in two-point crossover, two places are chosen
randomly then everything between these places is swapped between
parents. Then offsprings are passed to mutation stage which is a
genetic operation that alters some places of each chromosome at a
very low probability of occurrence denoted by $P_{m}$. Offsprings
create a new generation and this procedure continues until stop
condition has been reached, e.g., a solution is found or the user
specified maximum number of generations is evolved.

\vspace{1cm} \setcounter{section}{0}
 \setcounter{equation}{0}
 \renewcommand{\theequation}{I-\arabic{equation}}

{\Large{Appendix II}}\\
For the reader convenience we present here explicit realization of
generators $\lambda_{1},...,\lambda_{8}$ of Lie algebra $su(3)$:
$$
\lambda_{1}=\left(
\begin{array}{ccc}
  0 & 1 & 0 \\
  1 & 0 & 0 \\
  0 & 0 & 0 \\
\end{array}
\right)\quad ,\quad\lambda_{2}=\left(
\begin{array}{ccc}
  0 & -i & 0 \\
  i & 0 & 0 \\
  0 & 0 & 0 \\
\end{array}
\right)\quad ,\quad\lambda_{3}=\left(
\begin{array}{ccc}
  1 & 0 & 0 \\
  0 & -1 & 0 \\
  0 & 0 & 0 \\
\end{array}
\right)
$$
$$
\lambda_{4}=\left(
\begin{array}{ccc}
  0 & 0 & 1 \\
  0 & 0 & 0 \\
  1 & 0 & 0 \\
\end{array}
\right)\quad ,\quad\lambda_{5}=\left(
\begin{array}{ccc}
  0 & 0 & -i \\
  0 & 0 & 0 \\
  i & 0 & 0 \\
\end{array}
\right)\quad ,\quad\lambda_{6}=\left(
\begin{array}{ccc}
  0 & 0 & 0 \\
  0 & 0 & 1 \\
  0 & 1 & 0 \\
\end{array}
\right)
$$
$$
\lambda_{7}=\left(
\begin{array}{ccc}
  0 & 0 & 0 \\
  0 & 0 & 1 \\
  0 & 1 & 0 \\
\end{array}
\right)\quad ,\quad\lambda_{8}=\frac{1}{\sqrt{3}}\left(
\begin{array}{ccc}
  1 & 0 & 0 \\
  0 & 1 & 0 \\
  0 & 0 & -2 \\
\end{array}
\right).
$$

\newpage
{\bf Figure Captions}

{\bf Fig1:} Caption: $E(\rho)$ versus fidelity F for Werner state
$\rho_{_{W}}$. Parameters of program: $P_{m}=0.007$, $P_{c}=0.7$,
$N_{GA}=320$, $G_{GA}=300$, $N_{QN1}=400$ and $N_{QN2}=5$.
Resolution time for each generation is 1 seconds on a Celeron 2
GHz PC.

{\bf Fig2:} Caption: $E(\rho)$  versus q for GHZ-W mixture state
$\rho_{q}$. Parameters of program: $P_{m}=0.007$, $P_{c}=0.7$,
$N_{GA}=640$, $G_{GA}=300$, $N_{QN1}=500$ and $N_{QN2}=8$.
Resolution time for each generation is 3.75 seconds on a Celeron 2
GHz PC.

{\bf Fig3:} Caption: $E(\rho)$  versus $\alpha$ for two-qutrit
state $\rho_{_{\alpha}}$ . Parameters of program: $P_{m}=0.007$,
$P_{c}=0.7$, $N_{GA}=810$, $G_{GA}=300$, $N_{QN1}=800$ and
$N_{QN2}=10$. Resolution time for each generation is 45 seconds on
a Celeron 2 GHz PC.


\begin{thebibliography}{99}
\bibitem{nielsen1}
M. A. Nielsen and I. L. Chuang, Quantum Computation and Quantum
Information, Cambridge University Press, Cambridge, 2000.
\bibitem{ekert1}
 The Physics of Quantum Information: Quantum Cryptography,
Quantum Teleportation and Quantum Computation, edited by D.
Bouwmeester, A. Ekert, and A. Zeilinger, Springer, New York, 2000.
\bibitem{peres1}
A. Peres, Separability criterion for density matrices, Phys. Rev.
Lett. 77 (1996)  1413.
\bibitem{horod1}
M. Horodecki, P. Horodecki, and R. Horodecki, Separability
criterion and local information in separable states, Phys. Lett. A
223 (1996) 1.
\bibitem{wootters1}
W. K. Wootters, Entanglement of formation of an arbitrary state of
two qubits,  Phys. Rev. Lett. 80 (1998) 2245.
\bibitem{horod2}
P. Horodecki, M. Horodecki, and R. Horodecki, Bound entanglement
can be activated, Phys. Rev. Lett. 82 (1999) 1056.
\bibitem{lewen1}
B. Kraus, J. I. Cirac, S. Karnas, M. Lewenstein, Separability in
$2\times N$ composite quantum systems,  Phys. Rev. A 61 (2000)
062302.
\bibitem{nielsen2}
M. A. Nielsen and J. Kempe, Separable states are more disordered
globally than locally,  Phys. Rev. Lett. 86 (2001) 5184.
\bibitem{lewen2}
M. Lewenstein, B. Kraus, P. Horodecki, J. I. Cirac,
Characterization of separable states and entanglement witnesses,
Phys. Rev. A 63 (2001) 044304.
\bibitem{gurvitz}
L. Gurvits, Classical deterministic complexity of Edmonds' Problem
and quantum entanglement, in Proceedings of the 35th ACM Symposium
on the Theory of Computing, ACM Press, New York, 2003, pp. 10–-19.
\bibitem{toth1} G. T\'{o}th, and O. G\"{u}hne, Entanglement detection in the stabilizer formalism,
 Phys. Rev. A
72 (2005) 022340.
\bibitem{doherty1}
R. O. Vianna and A. C. Doherty, Distillability of Werner states
using entanglement witnesses and robust semidefinite programs,
Phys. Rev. A 74  (2006) 052306.
\bibitem{najar}
M. A. Jafarizadeh, G. Najarbashi, H. Habibian, Manipulating
multiqudit entanglement witnesses by using linear programming,
Phys. Rev. A 75 (2007) 052326.
\bibitem{bertl}
R. A. Bertlmann, H. Narnhofer and W. Thirring, Geometric picture
of entanglement and Bell inequalities, Phys. Rev. A 66 (2002)
032319.
\bibitem{branda1}
F. G. S. L. Brandao  and R. O. Vianna, Witnessed entanglement,
IJQI, 4  (2006) 331.
\bibitem{branda2}
F. G. S. L. Brandao, Quantifying entanglement with witness
operators, Phys. Rev. A 72  (2005) 022310.
\bibitem{mitchell}
M. Mitchell,  An Introduction to Genetic Algorithms, A Bradford
Book, The MIT Press, Cambridge, MA, 1999.
\bibitem{ramos1} R.V. Ramos, R.F. Souza, Calculation of the quantum entanglement measure of bipartite states, based on relative entropy, using
genetics algorithms, J. Comput. Phys. 175 (2002) 576.
\bibitem{ramos2} R. V. Ramos, Numerical algorithms for use in quantum
information, Journal of Computational Physics 192 (2003) 95.
\bibitem{vedral}
V. Vedral and M. B. Plenio, Entanglement measures and purification
procedures, Phys. Rev. A 57 (1998) 1619.
\bibitem{gisin1} N. Gisin, Hidden quantum nonlocality revealed by local filters, Phys. Lett. A 210 (1996) 151.
\bibitem{setinvar} Assuming the isometry of $\mathcal{E}$, it is easy to
show that $\mathcal{E}$ and subsequently $\mathcal{E}^\dag$ are
one to one and unto maps. By applyig $\mathcal{E}^\dag$  on the
both sides of the equality
$\mathcal{E}(\rho_{s})=\mathcal{E}(\sigma_{s})$ and using the
isometry condition we get $\rho_{s}=\sigma_{s}$ which implies
$\mathcal{E}$ is one to one. On the other hand for every
$\rho_{s}\in \mathcal{S}$ there exists  $\sigma_{s}\in
\mathcal{S}$ such that $\mathcal{E}(\sigma_{s})=\rho_{s}$, since
it is sufficient to take $\sigma_{s}=\mathcal{E}^\dag(\rho_{s})\in
\mathcal{S}$ which in turn imlies that $\mathcal{E}$ is an unto
map. Hence
$\mathcal{E}^\dag(\mathcal{S})=\mathcal{E}(\mathcal{S})=\mathcal{S}$.
\bibitem{hall1}
W. Hall,  Multipartite reduction criteria for separability, Phys.
Rev. A 72 (2005) 022311.
\bibitem{werner}
R. F. Werner, Quantum states with Einstein-Podolsky-Rosen
correlations admitting a hidden-variable model,  Phys. Rev. A 40
(1989) 4277.
\end{thebibliography}
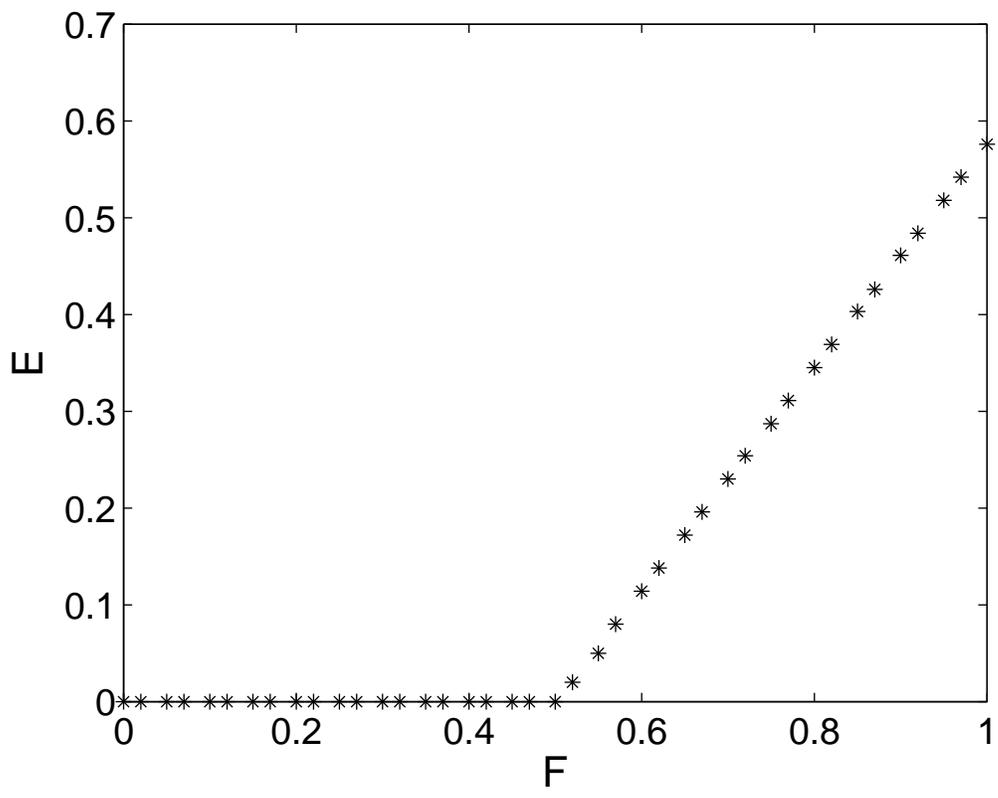
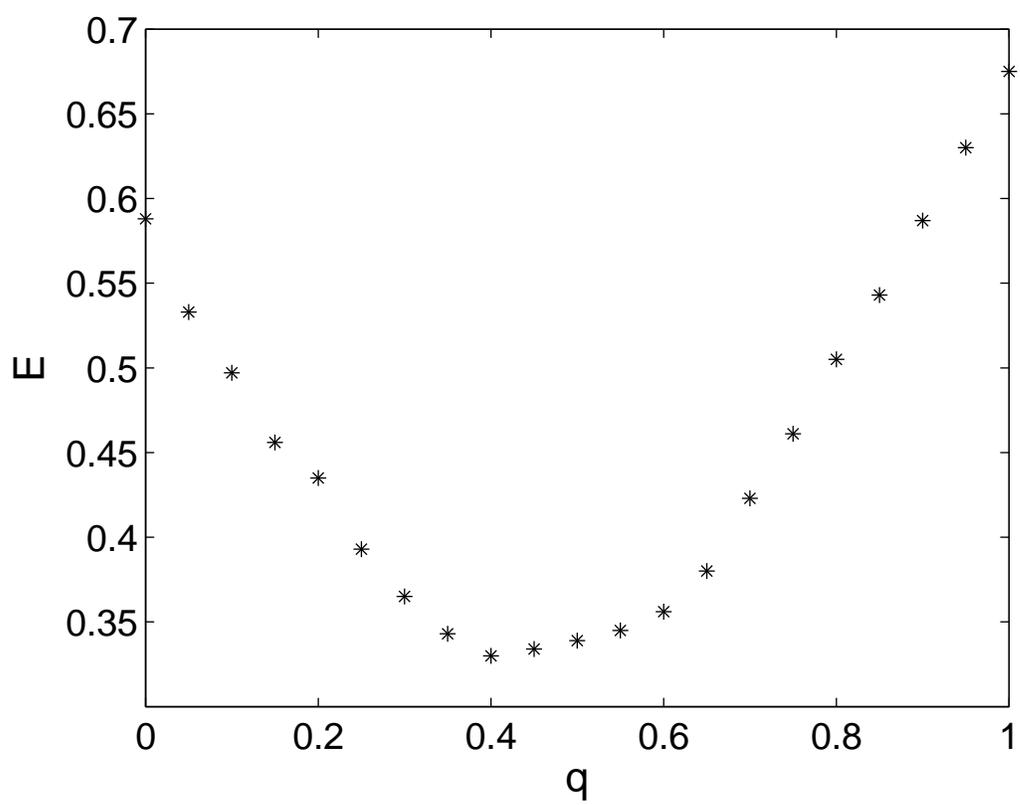
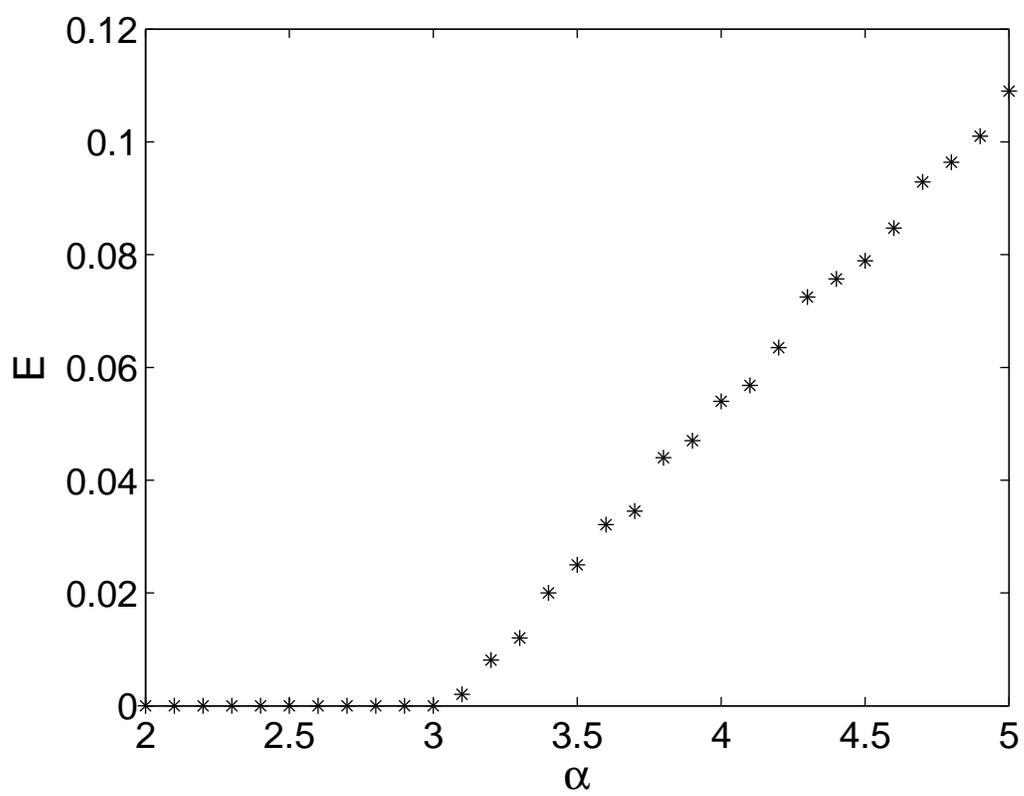
\end{document}